\documentclass[final,3p,times,twocolumn]{elsarticle}
 \biboptions{comma,sort&compress}
\usepackage{here}
\usepackage{amsfonts,amsmath,amssymb,bm,eucal}
\usepackage{graphicx,graphics}
\usepackage{epstopdf}
\usepackage{subfigure}

\def\nin{\noindent}
\def\beq{\begin{equation}}
\def\eeq{\end{equation}}
\def\bea{\begin{eqnarray}}
\def\eea{\end{eqnarray}}

\journal{Nuc. Phys. (Proc. Suppl.)}

\begin{document}

\begin{frontmatter}



 \title{
Holographic vector mesons in a hot and dense medium}


 \author{Floriana Giannuzzi}
\address{Istituto Nazionale di Fisica Nucleare, Sezione di Bari, 
\\
via Orabona 4,
 I-70126, Bari, Italy\\
Dipartimento di Fisica, Universit\`a degli Studi di Bari, 
via Orabona 4,
 I-70126, Bari, Italy}
\ead{floriana.giannuzzi@ba.infn.it}


\begin{abstract}
\noindent
We investigate vector meson spectral functions at finite temperature and density through the soft wall model, a bottom-up holographic approach to QCD. We find narrow resonances at small values of the parameters, becoming broader as temperature and density increase. We study dissociation of such states, occurring when no peak can be distinguished in the spectral function. We also find a decreasing of the mass of vector mesons at increasing temperature and density. Finally, a discussion of these results is presented.

\end{abstract}


\end{frontmatter}


\nin
The properties of vector mesons in a hot and dense medium can be studied by looking at their spectral functions. In fact, these objects carry information about the existence of bound states or resonances, their mass and width, and also about the properties of the medium, since transport coefficients and hydrodynamical quantities can be inferred from their low-energy behaviour.

The analysis of spectral functions at finite temperature and density is complicated, and still needs more hints; for a review of theoretical and experimental results, see \cite{Rapp:2010sj}. 

We compute spectral functions in a holographic bottom-up approach to QCD. Holographic means that it is inspired by the AdS/CFT correspondence, a conjecture put forward by Maldacena in 1997 \cite{maldacena1} stating that there is a duality between two different theories living in two different spacetimes: one is type IIB string theory living in a ten dimensional curved space, AdS$_5 \times S^5$, AdS$_5$ being a five-dimensional anti-de Sitter space and $S^5$ a five-dimensional sphere; the other theory is $\cal{N}$=4 Super Yang-Mills (SYM) theory living in a four-dimensional Minkowski space.
The metric of the AdS$_5$ space can be written through conformal coordinates:
\begin{equation}
ds^2=\frac{R^2}{z^2}(dt^2-d\bar x^2-dz^2)\,,
\end{equation}
where $R$ is the radius and $z$ the fifth coordinate.
The AdS space has a boundary, which is a Minkowski space and is defined by $z=0$.
The name {\it CFT} underlines the property of conformal invariance characterising the gauge theory.
The duality is accomplished by the identification of the coupling constants of the two theories and by the relation between the radius of the AdS space and the 't Hooft coupling of the gauge theory \cite{maldacena1}. 
Standing these two relations, a very important property of the correspondence follows: the supergravity limit of the string theory, which means very small coupling and very large radius, is dual to the limit of large $N_c$ and large 't Hooft coupling in the gauge theory.
In other words, the non-perturbative regime of the gauge theory is dual to a semiclassical theory in a curved spacetime.
For this reason, it can be interesting to study possible applications of such a conjecture to QCD, and try to compute some non-perturbative quantities through a perturbative approach. 
The models that have been introduced so far attempt to make the conjecture suitable for QCD, which, at odds with $\cal{N}$=4 SYM theory, is neither supersymmetric nor conformal.
In particular, a bottom-up approach will be followed, in which a five-dimensional theory is phenomenologically constructed, according to the dictionary established in \cite{gubser-klebanov-polyakov}: for each local gauge-invariant operator in the gauge theory there is a proper five-dimensional field, and its mass is related to the conformal dimension of the operator. 
Here the soft-wall model will be used \cite{sw}, in which conformal symmetry is broken by inserting a non-dynamical {\it dilaton-like} field in the action, $e^{-\phi}$, with $\phi=c^2z^2$. In this way, a mass scale, $c$, is introduced in the model, while the choice of $\phi$ is the simplest that can give Regge trajectories in hadron spectra. In fact, as an example, vector mesons can be analytically investigated, and, from the mass of the $\rho$ meson the parameter $c$ can be fixed: $c$=388 MeV.\\
We are interested in studying finite temperature and density effects.
In the five-dimensional theory this corresponds to modify the AdS space by adding a charged black-hole. 
Temperature and density are related to the position of the black-hole and its charge.
Moreover, a U(1) gauge field $A_0$ is introduced, dual to the QCD operator $\bar q\gamma_0 q$.
The resulting metric, known as the Reissner-Nordstr\"om (RN) metric, has the following expression:
\begin{equation}
ds^2=\frac{R^2}{z^2} \left(f(z)dt^2-d\bar x^2-\frac{dz^2}{f(z)}\right)
\end{equation}
\begin{equation}
f(z)=1-(1+Q^2)\left(\frac{z}{z_h}\right)^4+Q^2 \left(\frac{z}{z_h}\right)^6 \,,
\end{equation}
with  $0\leqslant Q\leqslant \sqrt{2}\,$, proportional to the black-hole charge \cite{Colangelo:2010pe}. The fifth coordinate $z$ is now bounded, $0<z<z_h$, $z_h$ being the outer horizon of the black hole, i.e. the lowest value of $z$ such that $f(z_h)=0$.

In this phenomenological model, we assume the following solution for $A_0$ \cite{Colangelo:2010pe}
\begin{equation}
A_0(z)=\mu - \kappa \frac{Q^2}{z_h^3} z^2\,,
\end{equation}
with $\kappa$ a parameter \cite{Colangelo:2010pe}, which verifies the condition required by the AdS/CFT correspondence that the source ($\mu$) of the operator ($\bar q\gamma_0 q$) must be equal to the boundary value of the five-dimensional field ($A_0(0)$). 
Temperature and density are linked to the black-hole parameters ($z_h$ and $Q$) in the following way. Temperature is defined by (Hawking temperature)
\begin{equation}
T=\frac{1}{4\pi} \left| \frac{df}{dz} \right|_{z_h} = \frac{1}{\pi z_h} \left( 1-\frac{Q^2}{2}\right)\,;
\end{equation}
the chemical potential is given by requiring a boundary condition on $A_0(z)$, stating that the field must vanish on the black-hole horizon ($ A_0(z_h)=0$), obtaining
\begin{equation}
\mu=\kappa\frac{Q}{z_h}\,.
\end{equation}
One can notice that both temperature and chemical potential are inverse functions of $z_h$, so the larger the space along the $z$-axis (black-hole horizon is far from the AdS boundary), the lower the temperature and chemical potential.

The objects we want to study through the model described so far are vector mesons. 
In the gauge theory, this corresponds to consider the local gauge-invariant operator $\bar q\gamma_\mu T^a q$ ($T^a$ are the Gell-Mann matrices); the field dual to such operator is a gauge vector field $V^a_M(x,z)$ described by the action
\begin{equation}\label{eqaction}
S=-\frac{1}{2\, k_V\, g_5^2}\int d^5x \sqrt{g}\, e^{-\phi(z)} \mbox{ Tr}\left[F_V^{MN}\, F_{V\, MN}\right]\,;
\end{equation}
$g$ is the determinant of the metric and $M,N$ are five-dimensional indices.
$F_V^{MN}=\partial^M V^N-\partial^N V^M$ is the strength tensor field (assuming a quadratic approximation for the action\footnote{the quadratic approximation is what we need for the computation of the spectral functions}).
The coefficient $k_V g_5^2=12\pi^2/N_c$ is fixed by a comparison of the perturbative terms of the two-point correlation functions in QCD and in the soft-wall model \cite{sw}.
We fix the gauge $V_z=0$.
Given the action \eqref{eqaction}, the Euler-Lagrange equation of motion for $V^i$ in the Fourier space reads
\begin{equation}\label{eomfield}
\partial_z\left( \frac{e^{-\phi(z)}}{z}f(z)\partial_z V_i(z,\omega^2)\right) + \frac{e^{-\phi(z)}}{z\, f(z)}\omega^2\, V_i(z,\omega^2)=0\,.
\end{equation}
The analysis of vector meson properties in a medium proceeds in two different ways, according to the values of temperature and density.
In case of very small values, a standard analysis, based on solving the equation of motion and looking for normalizable solutions, can be adopted, since the problem is very similar to the zero-density and zero-temperature case. On the other hand, when temperature and/or density become sizeable, a  different method, based on the computation of spectral functions, is needed.

\begin{figure}[h]
\centering
\includegraphics[width=7cm]{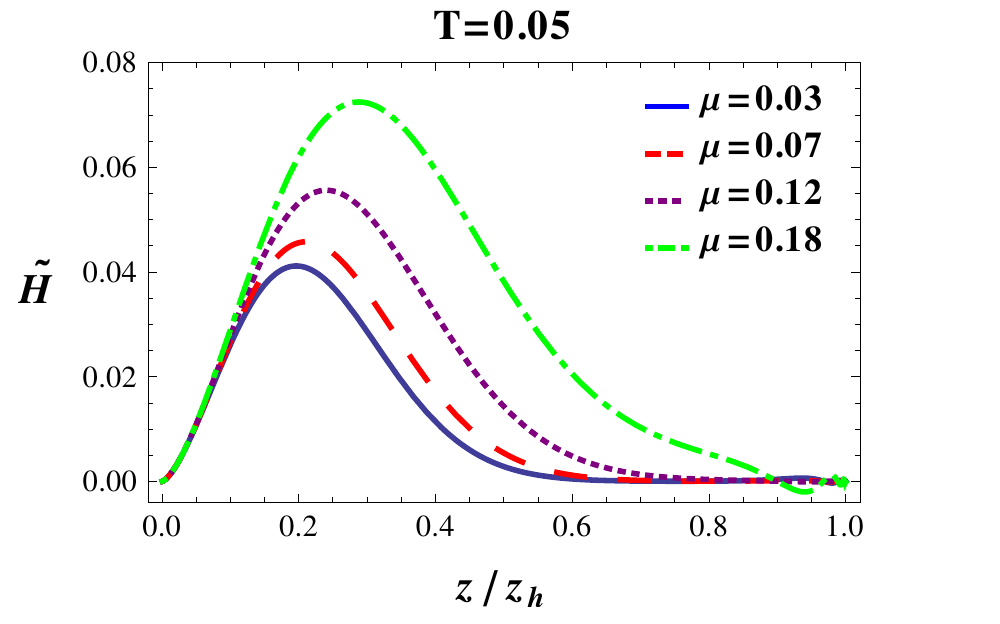}
\caption{Eigenfunctions of Eq. \eqref{scheq} for $T=0.05c$ and some values of $\mu$. $\kappa$=1 has been used.}
\label{fig:eigen}
\end{figure}

Let us first focus on the first case. As already pointed out, very small values of temperature and chemical potential correspond to very high values of $z_h$.
We perform a Bogoliubov transformation $V_i(z,\omega^2)=e^{B(z)/2} H(z,\omega^2)$, $B(z)=z^2+\log z-\log f(z)$, such that Eq.\eqref{eomfield} becomes
\begin{equation}\label{scheq}
 -\partial_z^2 H(z,\omega^2)+U(z)\, H(z,\omega^2)=\frac{\omega^2}{f(z)^2}\, H(z,\omega^2) 
\end{equation}
\begin{equation}
 U(z)=\frac{B'^2}{4}-\frac{B''}{2}\,.
 \end{equation}
We solve this Schr\"odinger-like equation, with boundary conditions $V^i(0)=0$ and $V^i$ normalizable, and find the eigenvalue, which is the mass of vector meson. 
This method can be used as long as $z_h$ is high enough that solutions of the equation of motion for the vector field vanish before reaching the black-hole horizon \cite{Colangelo:2012jy}.
Fig. \ref{fig:eigen} shows some examples of such solutions at $T$=0.05$c$ and some values of chemical potential (using $\kappa$=1); the green dash-dotted curve ($\mu=0.18$) represents a limit case, since $z_h$ is small enough to influence the solutions.
For higher values of chemical potential, this method cannot be used anymore.

\begin{figure}[h]
\centering
\includegraphics[width=7cm]{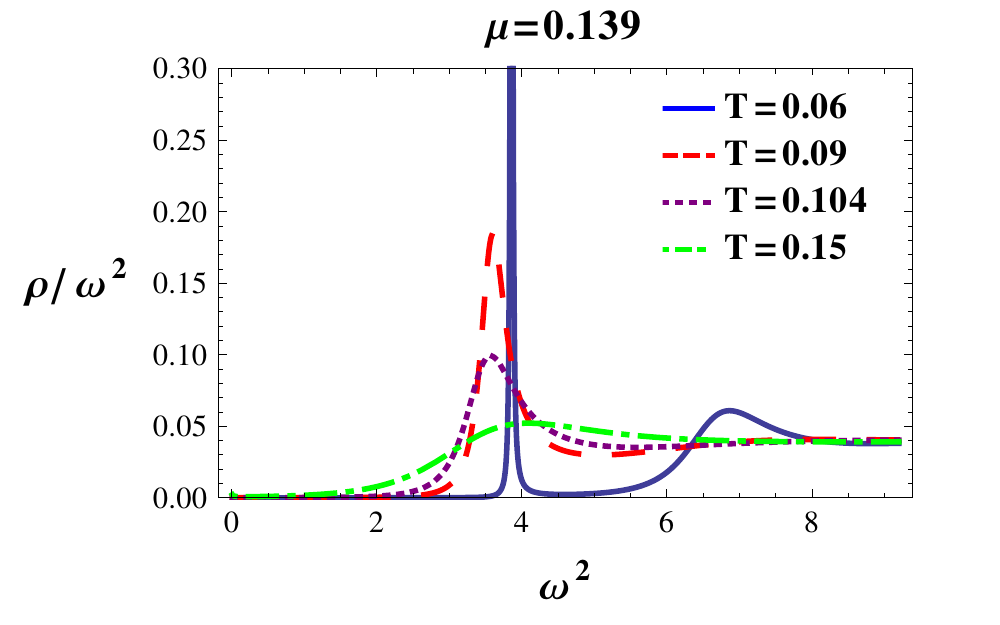}
\caption{Spectral functions of vector mesons for $\mu=0.139c$ and some values of $T$. $\kappa$=1 has been used.}
\label{fig:SF1}
\end{figure}

\begin{figure}[h]
\centering
\includegraphics[width=7cm]{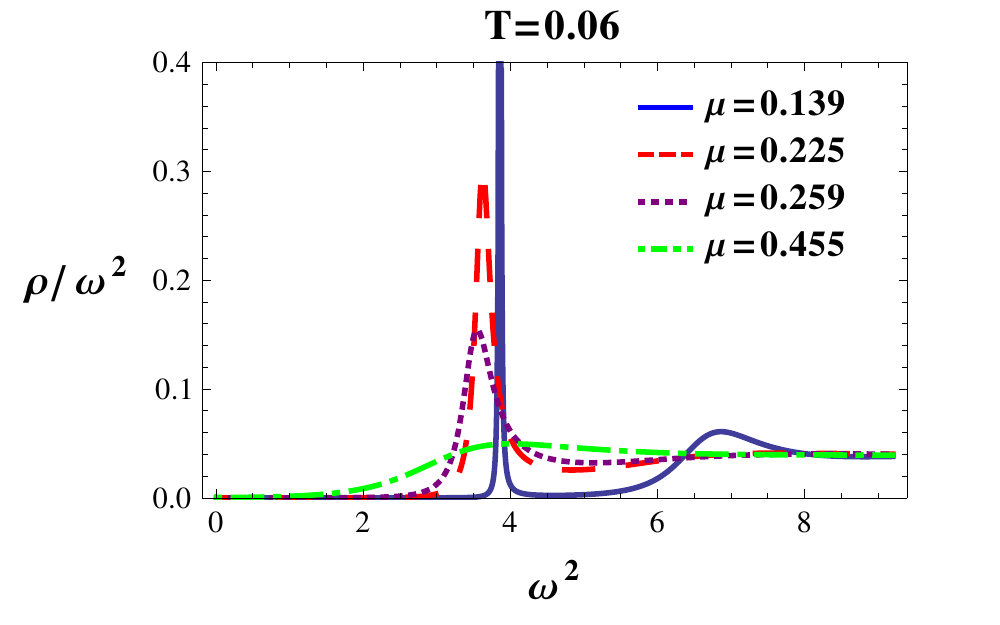}
\caption{Spectral functions of vector mesons for $T=0.06c$ and some values of $\mu$. $\kappa$=1 has been used.}
\label{fig:SF2}
\end{figure}

\begin{figure}[h]
\centering
\includegraphics[width=7cm]{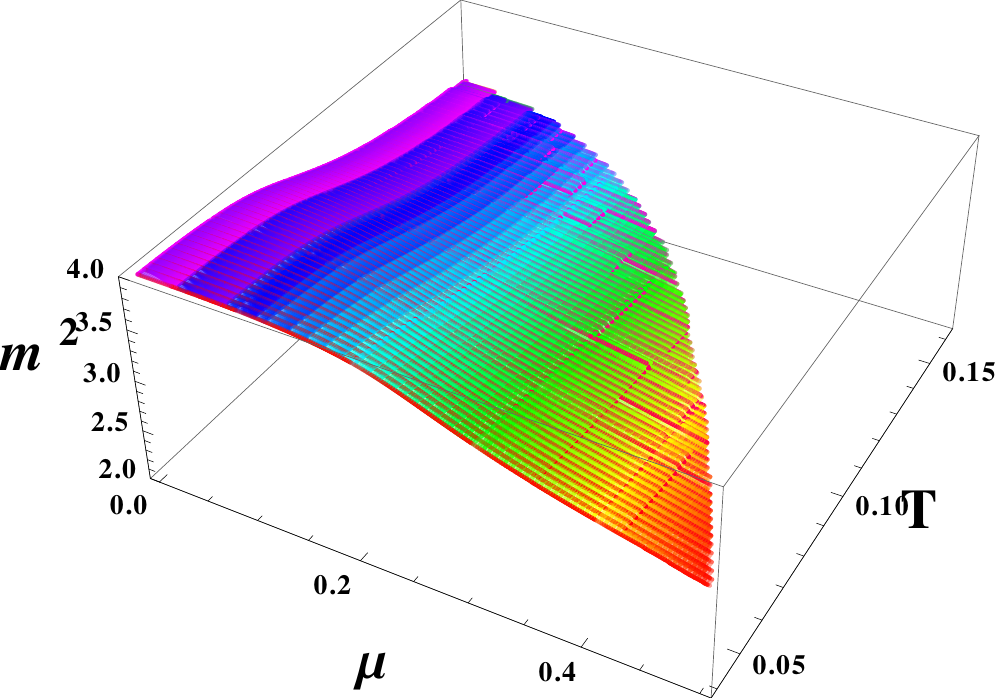}
\caption{Mass of the $\rho$ meson as a function of temperature and density. All quantities are in units of $c$; $\kappa$=1 has been used.}
\label{fig:mass}
\end{figure}

\begin{figure}[h]
\centering
\includegraphics[width=7cm]{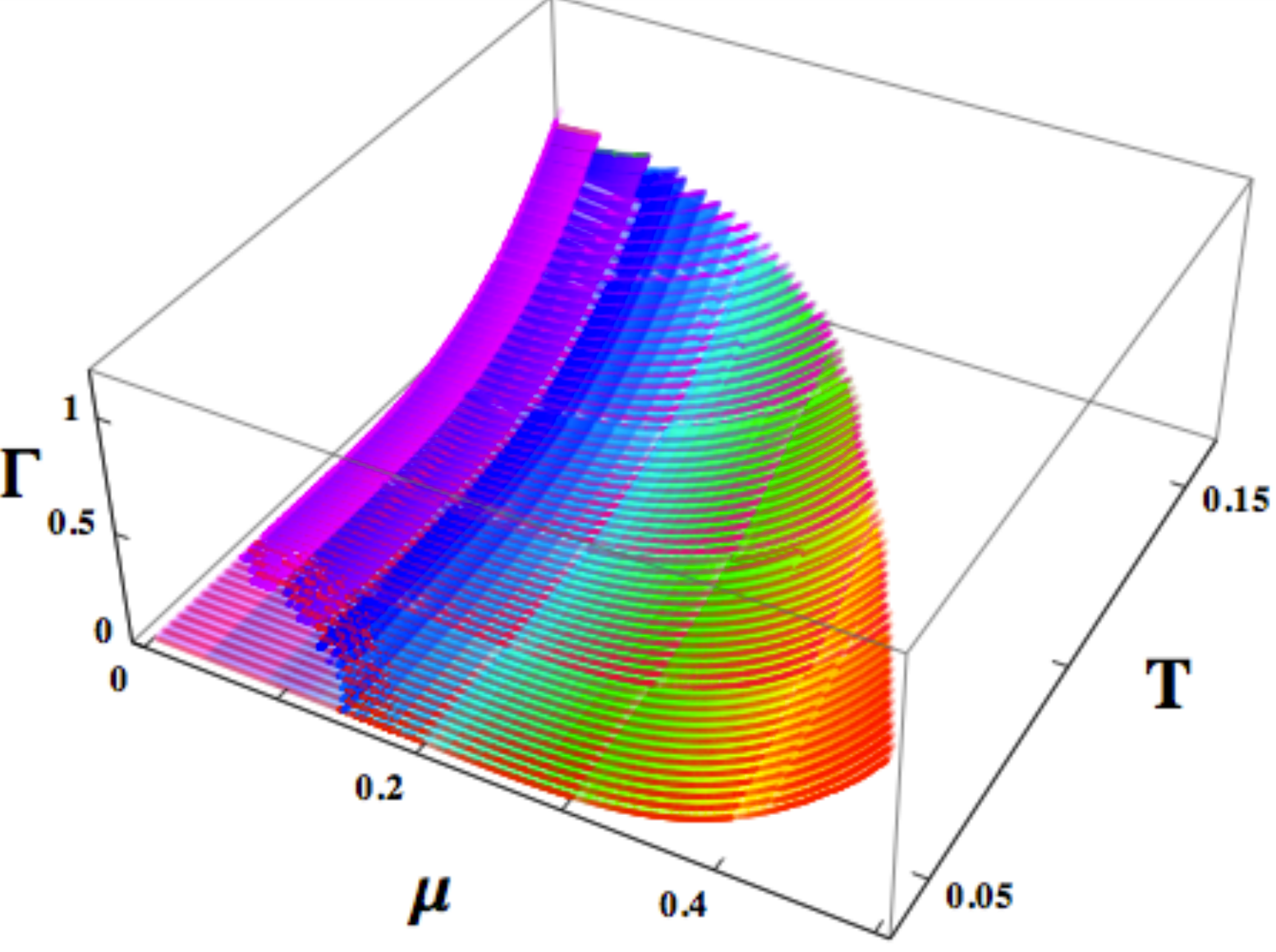}
\caption{Width of the $\rho$ meson as a function of temperature and density. All quantities are in units of $c$; $\kappa$=1 has been used.}
\label{fig:width}
\end{figure}

The other method consists in computing spectral functions, which are defined as the imaginary part of the retarded Green's function, which is in turn given by
\begin{eqnarray}
G^R_{ij}(\omega^2)&=&\frac{\delta^2 S}{\delta V_i^0(-\omega)\delta V_j^0(\omega)} \\
&=&\delta_{ij}\left. \frac{e^{-\phi(z)}\, f(z)}{g_5^2\, k_V} V(z,\omega^2)\frac{\partial_z\, V(z,\omega^2)}{z}\right|_{z=0}\,;   \nonumber
\end{eqnarray}
$V$ and $V_i^0$ are the bulk-to-boundary propagator and the source, respectively, of the vector field in the Fourier space, defined by $V_i(\omega^2,z)=V(z,\omega^2) V_i^0(\omega^2)$. $S$ is the action in \eqref{eqaction}.
The equation of motion for the bulk-to-boundary propagator is the same as for the vector field $V_i$ in \eqref{eomfield}, but the boundary conditions are that $V(0,\omega^2)$=1 and that near the black-hole horizon it must behave as the {\it falling in} solution:
\begin{equation}
V(z,\omega^2) \xrightarrow[]{z\to z_h} (1-z/z_h)^{-i\frac{\sqrt{\omega^2}\, z_h}{2(2-Q^2)}}\left(1+{\cal O}(1-z/z_h)\right)\,,
\end{equation}
a condition which ensures to get in the end the retarded Green's function, rather than the advanced one \cite{Policastro:2002se}.
In Figs. \ref{fig:SF1}-\ref{fig:SF2} some spectral functions are drawn, for $\mu$=0.139$c$ ($\kappa$=1) and some values of temperature, and for $T$=0.06$c$ and some values of $\mu$, respectively.
The position of the peaks of the spectral functions corresponds to the mass of resonances, in particular the first peak represents the ground state and the second one the first excited state.
The effect of raising temperature and/or density is to shift peaks towards small energies and make them broader, so the mass decreases and the width increases. This is evident in Figs. \ref{fig:mass}-\ref{fig:width}, where masses and widths are plotted versus temperature and chemical potential, in units of $c$. 
At small temperature and density the width is zero and the mass is the eigenvalue of Eq.\eqref{scheq}, while at higher values they are extracted from the spectral functions, by fitting each peak with a modified Breit-Wigner function
\begin{equation}
 \rho_{\mbox{\tiny BW}}(x)=\frac{a\, m\, \Gamma\, x^b}{(x-m^2)^2+m^2\Gamma^2}\,.
\end{equation}

An important feature of the spectral functions computed in this model is that they grow as $\omega^2$ at large energy, which is the expected behaviour \cite{Ding:2010ga}.
Moreover, they also show that excited states melt at lower temperature and desity with respect to the ground state, since the second peak disappears before the first.

In summary, the analysis of vector meson spectral functions in the soft wall model suggests that their width increases as temperature and density increase, meaning that these states become unstable, and their mass slightly decreases. The first result is exactly what we expected, and what other models and experiments find. For the second one, there is not a clear expectation, since many models and experiments support different outcomes. In fact, a dropping scenario has been predicted in \cite{Brown:1991kk}, while a mass-increase in \cite{Urban:1999im}. From the experimental point of view, data in \cite{Huber:2003pu} are compatible with a small increase of the mass, whereas the analyses in \cite{Arnaldi:2006jq} have found no significant effect. 
A wider and more quantitative discussion on these results can be found in \cite{Colangelo:2012jy}.

\nin

\section*{Acknowledgements}
\nin
I would like to thank P. Colangelo and S. Nicotri for collaboration.


\end{document}